\documentclass[aps,prl,preprint,groupedaddress]{revtex4-1}

\usepackage{natbib}
\usepackage{graphicx}

\begin{document}


\author{M.L. Reinle-Schmitt}
\email{philip.willmott@psi.ch}
\affiliation{Paul Scherrer Institut, 5232 Villigen, Switzerland}

\author{C. Cancellieri}
\affiliation{Paul Scherrer Institut, 5232 Villigen, Switzerland}

\author{S.J. Leake}
\affiliation{Paul Scherrer Institut, 5232 Villigen, Switzerland}

\author{A. Cavallaro}
\affiliation{Dept. of Materials, Imperial College London, London SW7 2AZ, UK}

\author{J.A. Kilner}
\affiliation{Dept. of Materials, Imperial College London, London SW7 2AZ, UK}

\author{E. Pomjakushina}
\affiliation{Paul Scherrer Institut, 5232 Villigen, Switzerland}

\author{P.R. Willmott}
\affiliation{Paul Scherrer Institut, 5232 Villigen, Switzerland}

\date{\today}

\title{Chemistry and structure of homoepitaxial SrTiO$_3$ films and their influence on oxide-heterostructure interfaces}

\begin{abstract}
The properties of single-crystal SrTiO$_{3}$ substrates and homoepitaxial SrTiO$_{3}$ films grown by pulsed laser deposition have been compared, in order to understand the loss of interfacial conductivity when more than a critical thickness of nominally homoepitaxial SrTiO$_{3}$ is inserted between a LaAlO$_{3}$ film and a SrTiO$_{3}$ substrate. In particular, the chemical composition and the structure of homoepitaxial SrTiO$_{3}$ investigated by low-energy ion-scattering and surface x-ray diffraction, show that for insulating heterointerfaces, a Sr-excess is present between the LaAlO$_{3}$ and homoepitaxial SrTiO$_{3}$. Furthermore, an increase in the out-of-plane lattice constant is observed in LaAlO$_{3}$, indicating that the conductivity both with and without insertion of SrTiO$_{3}$ thin film originates from a Zener breakdown associated with the polar catastrophe. When more than a critical thickness of homoepitaxial SrTiO$_{3}$ is inserted between LaAlO$_3$ and SrTiO$_3$, the electrons transferred by the electronic reconstruction are trapped by the formation of a Sr-rich secondary phase and Sr-vacancies. The migration of Sr towards the surface of homoepitaxial STO and accompanying loss of interfacial conductivity can be delayed by reducing the Sr-content in the PLD target.
\end{abstract}

\maketitle

Complex-oxide-heterostructures have been the subject of intense research in recent years. A prominent example is the interface between LaAlO$_{3}$ (LAO) and SrTiO$_{3}$ (STO) --- despite both components being band insulators, a two-dimensional electron system is created at their interface \cite{Ohtomo2004n} after deposition of a critical LAO thickness of four monolayers (MLs) \cite{Thiel2006s}. Phenomena such as magnetoresistance oscillations \cite{Caviglia2010prl}, superconductivity \cite{Reyren2007s}, carrier density tunability \cite{Thiel2006s}, and atomic force microscopy (AFM) writing \cite{Cen2009s} make this system technologically interesting for future integrated oxide-based electronic devices. However, extensions of this phenomenon crucial to future applications, in particular its observation on other substrate types and the successful fabrication of multilayer LAO/STO containing multiple conducting layers, appear to be thwarted by material issues \cite{Chakhalian2012nm}, mostly regarding interface quality --- related structures such as LAO on homoepitaxial STO (h-STO), LAO on a STO buffer layer to other substrate types, and LAO/STO superlattices, have been grown under a wide variety of growth and annealing conditions \cite{Brinks2011apl, Hernandez2012prb, Bark2011pnac, Gunkel2012apl, Park2010nc}. The large variation in the reported electronic properties is indicative of how sensitive this system is to growth parameters. For example, it was reported that the interfacial carrier density of heterostructures of $15$~ML LAO on $N$~ML h-STO ($3 \leq N \leq 6$) decrease with increasing $N$ \cite{Fix2009apl}, raising questions on the properties of ultrathin h-STO films compared to single-crystal STO substrates. Therefore, understanding the electronic properties of samples grown on buffer h-STO is key towards successful fabrication of superlattices with multiple conducting interfaces. 

In the case of STO single crystals, local deviations from the expected surface chemistry of TiO$_{2}$-terminated STO were found to be induced by Sr-migration towards the surface \cite{Desu1990jacs} or desorption of titanium \cite{Nishihara1997mrssp}, depending on the processing parameters, while other phase- and structural changes have also been observed \cite{Herger2007prl, Tokuda2011apl, Yoshimoto1994apl}. In addition, global off-stoichiometry and formation of oxygen and cations vacancies can occur as a result of noncongruent transfer from the PLD target to the growing film \cite{Ohnishi2005apl,Kozuka2010apl}. 

In the following, we report on the investigation of the change in interfacial properties when LAO is separated from the bulk substrate by an ultrathin layer of h-STO, and why, when more than a critical thickness of h-STO is grown, the conductivity disappears entirely. Low-energy ion-scattering (LEIS) was used to study the near-surface chemical of h-STO and to show that a Sr-excess builds up with increasing h-STO thickness. Changes in the out-of-plane lattice parameter of LAO with h-STO thickness was observed by surface x-ray diffraction (SXRD), confirming charge transfer across the interface \cite{Pauli2011prl,Cancellieri2011prl}, independent of the h-STO thickness. It is proposed, however, that above the critical h-STO thickness, these itinerant electrons become trapped in the increasing number of vacancies formed in the h-STO layer. Moreover, we demonstrate that this defect density can be reduced by changing the stoichiometry of the STO PLD target, thereby increasing the critical h-STO thickness.

\begin{figure}
\centering
\includegraphics[width=8cm]{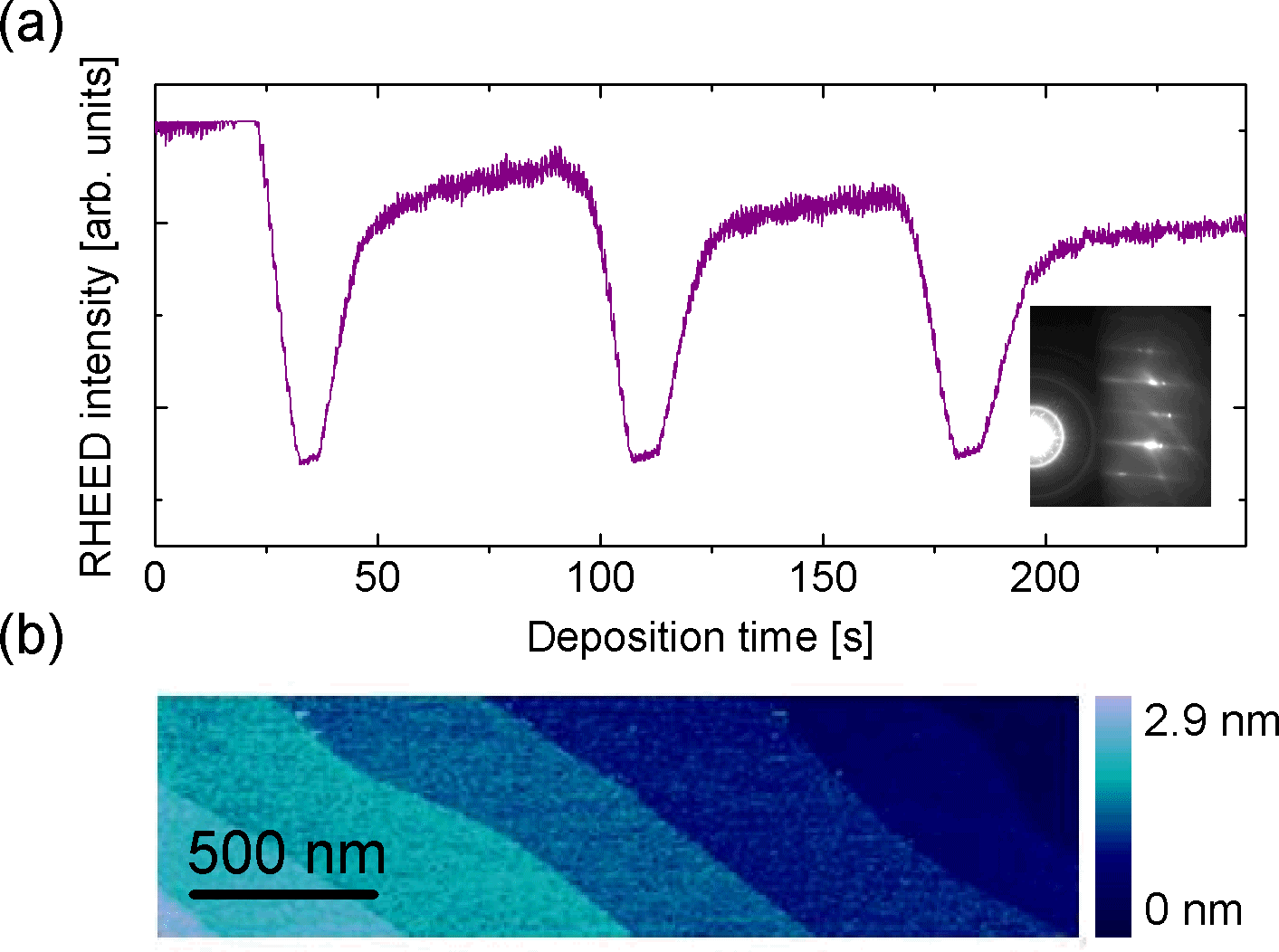}
\caption{(a) Modulation of the specular RHEED intensity during deposition and experimental RHEED pattern (inset) from the surface of a $3$~ML-thick h-STO grown with the ``standard'' STO target on TiO$_{2}$-terminated STO substrate. (b) AFM image of $5$~ML-thick h-STO on a TiO$_{2}$-terminated STO substrate}
\label{fig:figure1}
\end{figure}

Bilayers of LAO on h-STO were grown by PLD on TiO$_{2}$-terminated STO substrates (miscut $< 0.1^{\circ}$). The substrates were heated to $650^\circ$C and growth of both h-STO and LAO films was performed at an oxygen pressure of $8 \times 10^{-5}$~mbar, a laser fluence of $0.5$~J$\cdot$cm$^{-2}$, and repetition rate of $1$~Hz. The growth rate was monitored by reflection high-energy electron-diffraction [RHEED, see e.g., Fig.~\ref{fig:figure1}(a) for the growth of $3$~ML h-STO]. The deposition was paused after each ML of h-STO for $\approx 40$~s in order to allow the surface to relax (observed by a levelling out in the RHEED intensity) \cite{Blank2000jcc}. After growth, the temperature was decreased to $550^\circ$C and the films were post-annealed for $1$~hour in $200$~mbar O$_2$ in order to remove oxygen vacancies. Two PLD targets were used for the h-STO growth: a stoichiometric STO single crystal, referred to henceforth as the ``standard'' target; and a $3$~\%\ Sr-deficient polycrystalline target (Sr$_{0.97}$TiO$_{3 - \delta}$, density $> 85$~\%). The spotty 2-D RHEED pattern, including Kikuchi lines, after h-STO growth [see inset of Fig.~\ref{fig:figure1}(a)] demonstrates both layer-by-layer growth and high crystalline quality. The AFM image of $5$~ML h-STO in Fig.~\ref{fig:figure1}(b) shows atomically flat terraces. The h-STO layers were studied by LEIS at the Department of Materials, Imperial College, London. Prior to measurement, the samples were cleaned with ethanol, acetone and millipore water, and then {\em in-situ} by an oxygen atomic plasma for $30$~minutes. LEIS was performed using $3$~keV $^{4}$He$^{+}$ ions. Etching for depth-profiling was achieved using $0.5$-keV $^{40}$Ar$^{+}$ ions. SXRD measurements were performed at the Material Science beamline, Swiss Light Source \cite{Willmott2013jsr} at a photon energy of $15.5$~keV.

\begin{figure}
\centering
\includegraphics[width=9cm]{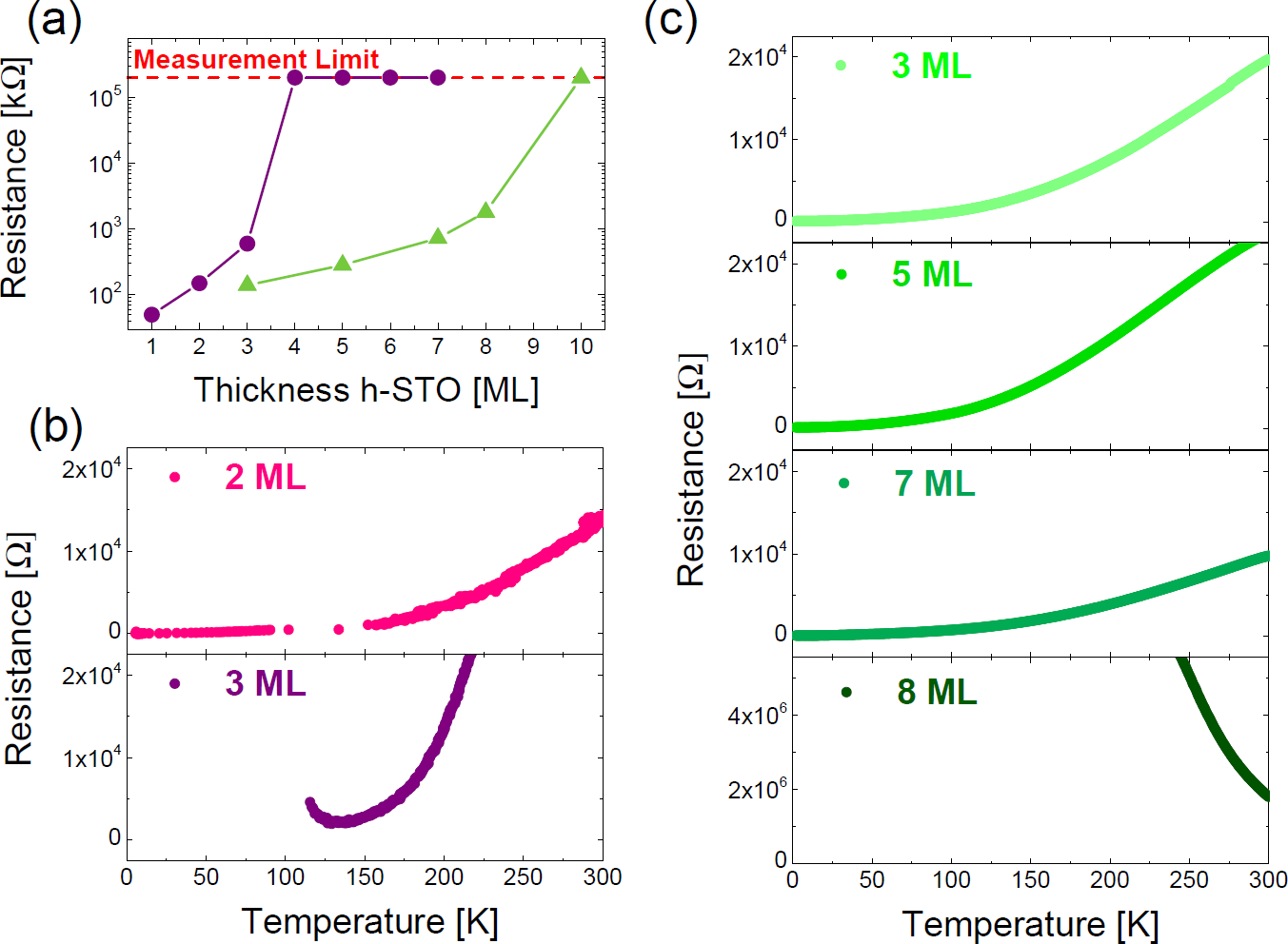}
\caption{(a) Dependence of the room-temperature resistance of $5$~ML LAO/h-STO/STO-substrate heterostructures on the thickness of the h-STO layer, for ``standard'' (violet circles) and Sr-deficient STO (green triangles) PLD targets. Resistance versus temperature of heterostructures grown with (b) the ``standard'' STO target, and (c) the Sr-deficient STO target.}
\label{fig:figure2}
\end{figure} 

\begin{figure*}
\centering
\includegraphics[width=16cm]{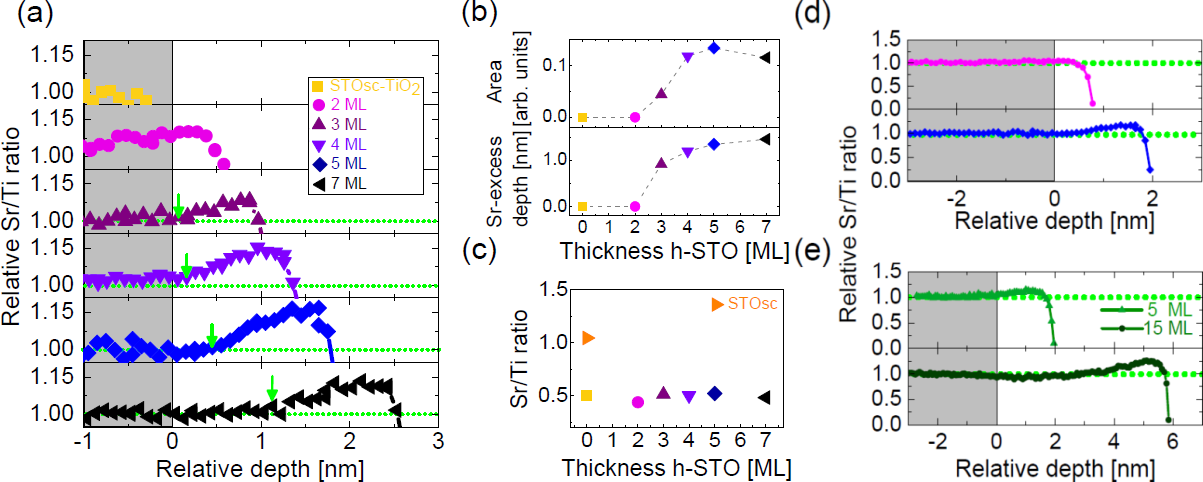}
\caption{(a) Depth profile across the interface of the Sr/Ti ratio for different thicknesses of h-STO films grown with the ``standard'' target, two bare single-crystal substrates, one mixed-terminated (STOsc), and one TiO$_{2}$-terminated (STOsc-TiO$_2$). The onset of the Sr-rich region is indicated with a green arrow. (b) ``Standard'' h-STO thickness dependence of the area below the Sr/Ti ratio curve and extent of the Sr-rich region. (c) ``Standard'' h-STO thickness dependence of the surface Sr/Ti ratio of the same samples.  The depth profile across the interface of the Sr/Ti ratio for thicknesses of h-STO films below and above the critical thickness grown with the ``standard'' (d) and Sr-deficient (e) targets, a linear background was subtracted from the data of the $2$~ML ``standard'' and $5$~ML Sr-deficient h-STO samples}
\label{fig:figure3}
\end{figure*}

Conductivity measurements revealed that all heterostructures with either $> 3$~ML ``standard'' h-STO or $> 8$~ML Sr-deficient h-STO are insulating ($N_{LAO}\geq 4$). The resistance of both types of interface versus h-STO thickness is shown in Fig.~\ref{fig:figure2}(a). The conducting samples have sheet resistances in a similar range as that for standard LAO on STO \cite{Thiel2006s}. Indeed, this excellent agreement is a strong argument that there is no significant contribution from oxygen vacancies in the h-STO films. All interfaces with less than $4$~ML LAO are insulating. The interfacial resistance measured as a function of temperature shows an upturn with increasing h-STO thickness for both types of target stoichiometries [Fig.~\ref{fig:figure2}(b) and (c)], explained by charge localization (i.e., freezing of the carriers) due to point defects in the h-STO. For the films grown using the Sr-deficient target, this upturn appears at a larger h-STO film thickness, indicating that a decrease in the target's strontium content delays the formation of defects during the PLD growth process. This raises the issue of which type of defect leads to this behaviour.  

Under our growth conditions, cation-vacancy formation is more likely than oxygen vacancies \cite{Kozuka2010apl, Meyer2002joe}, because, on the one hand, the films are post-annealed in oxygen, and, on the other, nonstoichiometric ablation of the target can occur. First-principle calculations from Tanaka \textit{et al.} \cite{Tanaka2003prb} indicate that Sr-vacancy formation is energetically favored over Ti-vacancy formation. This begs the question of where the Sr-species relocate after producing the vacancies. Phase separation involving Sr has already been reported in the literature both in thin films and single crystals \cite{Jung2012ees, Meyer2002joe, Tokuda2011apl, Szot2000ss, Meyer2003prl, Kim2009apl, Hellberg2012prl}. Two main mechanisms were proposed for the formation of Sr-vacancies. Firstly, partial Sr-Schottky defects can be formed in the film, thereby creating oxygen- and Sr-vacancies, whereby the O and Sr create ordered superlattices of Ruddlesden-Popper structures (${\rm SrO})_{\rm RP}$ \cite{Moos1997jacs}, as experimentally observed in Sr-rich h-STO thin films \cite{Tokuda2011apl}:
\begin{equation}
{\rm Sr}_{\rm Sr}^{x} + {\rm O}_{\rm O}^{x}  \rightleftharpoons {\rm V_{\rm Sr}^{''}} + {\rm V_{\rm O}^{..}}+({\rm SrO})_{\rm RP},
\end{equation}  
whereby V$_{\rm Sr}^{''}$ is a Sr-vacancy and V$_{\rm O}^{..}$ is an O-vacancy, according to the formalism of Kr\"oger and Vink \cite{KV}.
Secondly, Sr and O can react to form a separate Sr-rich phase specifically at the surface of h-STO, leaving doubly charged Sr-vacancies \cite{Meyer2002joe, Gunkel2012apl}:
\begin{equation}
{\rm Sr}_{\rm Sr}^{x} + \frac{1}{2} {\rm O}_{2} + 2 e^- \rightleftharpoons \rm V_{\rm Sr}^{''} + ({\rm SrO})_{\rm sp}
\end{equation} 
whereby V$_{\rm Sr}^{''}$ is a doubly charged Sr-vacancy and SrO$_{\rm sp}$ is the secondary phase.

In Fig.~\ref{fig:figure2} it is shown that the h-STO thickness threshold to suppress conductivity can be tuned by decreasing the Sr-content in the PLD target. This suggests that the film stoichiometry influences the electron-transfer process. Furthermore, both of the above explanations for the accumulation of defects involve a second-phase formation. LEIS measurements were therefore performed in order to investigate the h-STO stoichiometry. This technique is highly surface sensitive \cite{Brongersma2006ssr}. The stoichiometry depth profiles of different films were thus accurately determined by LEIS combined with controlled etching.

The LEIS/depth profiling results are summarized in Fig.~\ref{fig:figure3}(a) for several ``standard'' h-STO films, plus bare substrates for comparison. The Sr/Ti ratio increases in the near-surface region of approximately the first one to two monolayers for films thicker than $3$~ML. However, below the critical thickness of $2$~ML, the Sr/Ti ratio of the h-STO is the same as the bulk ratio to within $3.5$~\%. The depth of the top region with altered stoichiometry shown in Fig.~\ref{fig:figure3}(b) increases slightly with h-STO film thickness after the initial threshold at $3$~ML and becomes constant at approximately $1.5$~nm for h-STO thicknesses above $5$~ML. In addition, the initial Sr/Ti ratio at the surface was investigated [Fig.~\ref{fig:figure3}(c)] for the h-STO films, plus mixed-terminated and TiO$_{2}$-terminated STO single crystals. The Sr/Ti ratio for the latter is around $0.5$, which compares well with the ratio found at the surface of all the h-STO films, from which we can conclude that they maintain the TiO$_{2}$-termination after growth, at least for the range of thicknesses studied. The Sr/Ti ratio across ``standard'' h-STO films is compared with that of Sr-deficient h-STO films with thicknesses below and larger than the critical thickness in Fig.~\ref{fig:figure3}(d) and (e). The films thinner than the critical thickness derived from the conductivity measurements ($2$~ML ``standard'' h-STO and $5$~ML Sr-deficient h-STO) exhibit very limited Sr-excess at the surface. By increasing the number of deposited layers, the Sr-excess builds up at the surface of both ``standard'' and Sr-deficient STO. This result proves that the Sr-migration is delayed by decreasing the Sr-content in the PLD target, similar to results obtained by Jung \textit{et al.} \cite{Jung2012ees}.

\begin{figure}
\centering
\includegraphics[width=6cm]{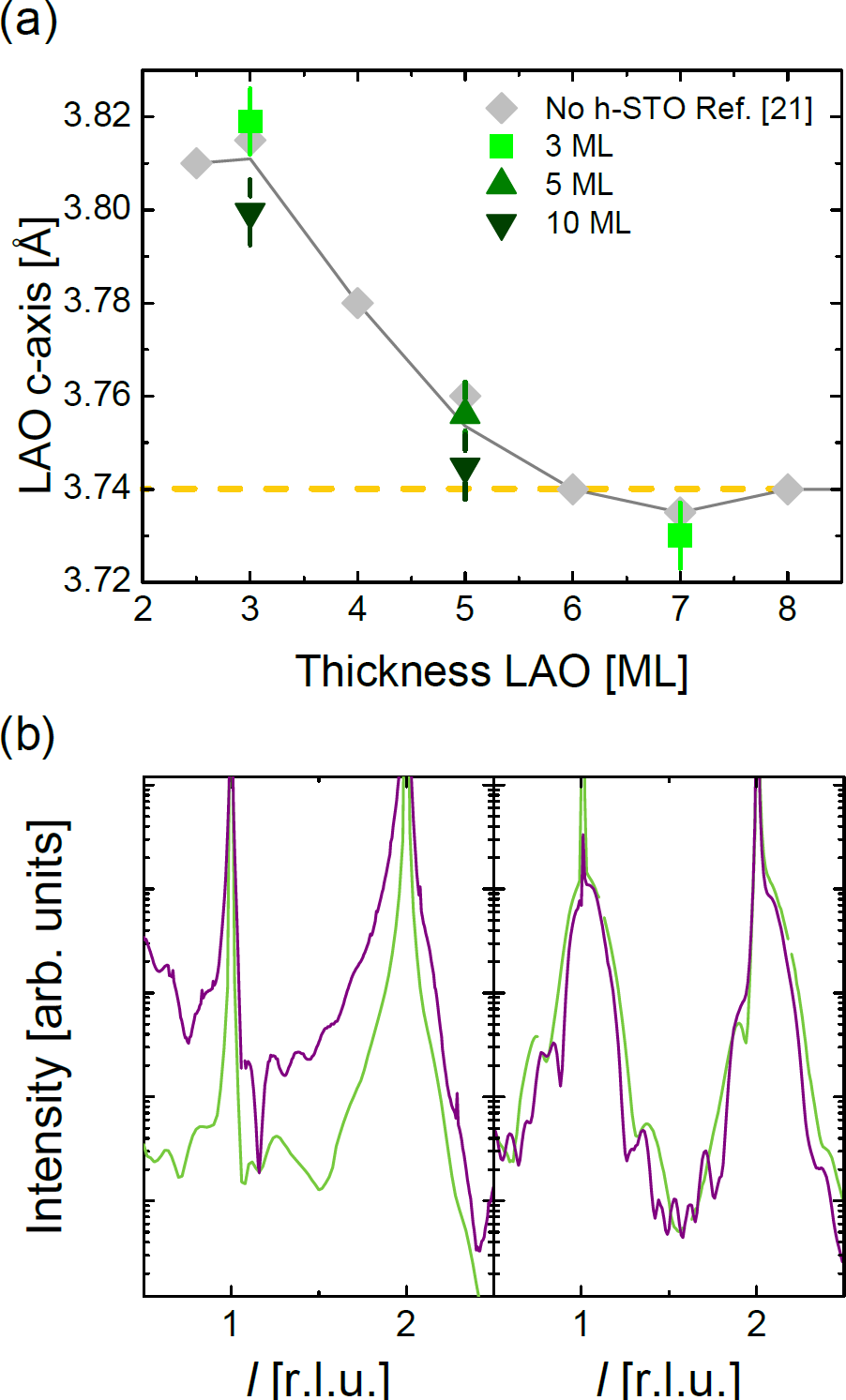}
\caption{(a) Out-of-plane lattice parameter of LAO/h-STO heterostructures obtained by simulation of the specular CTR scans (along with their error bars), the grey line is a guide to the eye and the dashed orange line corresponds to the elastic limit of the strained out-of-plane lattice parameter. (b) Intensity of the specular CTR scans of (left): $5$~ML h-STO grown with the ``standard'' target in violet and Sr-deficient target in green, (right): $5$~ML LAO on top of h-STO grown with the ``standard'' and Sr-deficient targets, as a function of $l$ in STO reciprocal lattice units [r.l.u.].}
\label{fig:figure4}
\end{figure}

The strong correlation between Fig.~\ref{fig:figure2}(a) and Figs.~\ref{fig:figure3}(b), (c) and (d) suggests that the onset of insulating behaviour of the interface is related to the Sr-rich layer. To explain a mechanism responsible for the loss of conductivity after a given h-STO thickness, the origin of the conductivity in samples with thinner h-STO interlayers needs to be known. Because the onset of conductivity as a function of LAO thickness is identical in both the standard LAO/STO system and the LAO/h-STO/STO systems, it seems probable that the polar catastrophe is also responsible for conductivity at interfaces involving h-STO \cite{Thiel2006s, ReinleSchmitt2012nc}. A signature of the intrinsic polar-catastrophe mechanism is a dilation of the LAO out-of-plane lattice parameter, due to one or both phenomena of the increased ionic radius of trivalent titanium associated with interfacial intermixing \cite{Willmott2007prl, Pauli2011prl} and electrostriction in films in which the electric field is still present \cite{Cancellieri2011prl}. SXRD measurements were performed in order to extract the out-of-plane lattice parameter of the LAO films. Simulations of the film peak position allows an accurate determination of the lattice parameter of LAO, shown in Fig.~\ref{fig:figure4}(a) for heterostructures both with and without the h-STO interlayer. Samples with $3$~ML LAO exhibit, as expected, a larger out-of-plane lattice parameter of $\approx 3.80$~\AA. In contrast, the lattice parameter decreases to $3.74$~\AA\ for heterostructures with LAO thicknesses greater or equal to $4$~ML. From these measurements, we conclude that the increase of the lattice parameter in LAO remains when h-STO is inserted between LAO and the STO substrate \cite{Cancellieri2011prl} and that the polar catastrophe is responsible for the formation of the interfacial electron gas.

Importantly, samples with a sufficient h-STO thickness to induce an insulating interface show a similar change in the out-of-plane lattice parameter. It appears that the Zener breakdown occurs, but the transferred electrons are trapped. Bulk conductivity in STO is known to be affected by extrinsic and intrinsic doping \cite{Moos1997jacs,Gunkel2012apl}. Intrinsic self-dopants under our growth/annealing conditions are far more likely to be acceptor-type Sr-vacancies than donor-type oxygen vacancies. Therefore, the loss of conductivity after a critical h-STO thickness would imply that the number of Sr-vacancies becomes sufficient to trap the transferred electrons, and would also explain the fairly gradual loss of conductivity with h-STO thickness. By trapping the electrons in Sr-vacancies, an ionic self-compensation mechanism \cite{Meyer2002joe} takes place. A similar example of carrier deactivation in doped STO has already been reported when electrons supplied by extrinsic Nb-dopants are compensated by Sr-vacancies introduced during growth \cite{Kozuka2010apl, Hellberg2012prl}. 

The increase in the critical h-STO thickness needed to suppress conductivity when using a Sr-deficient PLD target can be explained by the well-known problem of noncongruent transfer in PLD \cite{Willmott2000rmp,Willmott2004pss}. If a stoichiometric STO target produces Sr-rich layers, then presumably marginally lowering the Sr-content of the target will compensate this to an extent. Indeed, this change in the electron density is visible from the Laue fringes in the specular CTRs of $5$~ML of h-STO grown with the ``standard'' and Sr-deficient targets [Fig.~\ref{fig:figure4}(b)], which are significantly more pronounced in the ``standard'' sample. Unfortunately, the Sr-content can only be decreased by a few percent before separate STO and TiO$_2$ phases are visible in the diffraction pattern.

In conclusion, we have found that the interface between LAO and h-STO becomes insulating when h-STO exceeds a given critical thickness, which depends on the Sr-content of the PLD target and subsequent segregation of Sr to the top of the h-STO layer. For h-STO films close to the critical thickness, the Sr-rich layer region coincides with an upturn in conductivity at low temperature, due to defects in the interface between LAO and h-STO. The simultaneous formation of Sr-vacancies and a Sr-rich secondary phase is predicted by theory \cite{Moos1997jacs}. SXRD measurements of the LAO out-of-plane lattice parameters of LAO on h-STO heterostructures support the polar-catastrophe scenario as the origin of the conductivity for interfaces grown on h-STO. An ionic self-compensation mechanism, in which electrons are trapped in acceptor-type Sr-vacancies, explains the loss of conductivity above a critical number of these defects, which is delayed to thicker h-STO thicknesses in films grown using a Sr-deficient PLD target. This work provides a new insight into the non-conducting behaviour observed when LAO is grown on h-STO and provides a basis for future works on superlattices with multiple conducting interfaces.

\section*{Acknowledgments}

Support of this work by the Schweizerischer Nationalfonds zur F\"{o}rderung der wissenschaftlichen Forschung, in particular the National Center of Competence in Research, Materials with Novel Electronic Properties, MaNEP is gratefully acknowledged. One of the authors (A.C.) acknowledges the financial support of the European Union (FP7-PEOPLE-2010-IEF, Project number 274730). The authors would like to thank Mr George Harrington for his help for the LEIS measurements, Dr C. W. Schneider for useful discussions and Dr M. Medarde for PPMS measurements.

%

\end{document}